\documentclass[10pt]{article}
\begin{document}
\title{Reply to Comment by Galapon on 'Almost-periodic time observables for bound quantum systems'}
\author{Michael J. W. Hall\\
Theoretical Physics, IAS, \\ Australian National
University,\\
Canberra ACT 0200, Australia}
\date{}
\maketitle



\begin{abstract}
In a recent paper \cite{hall}, I made several critical remarks on a `Hermitian time operator' proposed by Galapon \cite{galapon}. Galapon has correctly pointed out that remarks pertaining to `denseness' of the commutator domain are wrong \cite{comm}.  However, the other remarks still apply, and it is further noted that a given quantum system can be a member of this domain only at a set of times of total measure zero.
\end{abstract}

It has been known for some decades that a canonical time observable $T_{can}$ exists for quantum systems having a continuous energy spectrum, such as a free particle \cite{kij,holevo}, and for periodic quantum systems such as a harmonic oscillator \cite{helstrom}.  The canonical time observable is described by a probability operator measure (POM), and the corresponding  probability density satisfies the fundamental covariance property
\begin{equation} \label{cov}
 p(t|\psi_\tau) = p(t-\tau|\psi_0),
\end{equation}
where $\psi_\tau$ denotes the wavefunction of the system at time $\tau$.  For a nondegenerate energy spectrum the canonical time probability density takes the form
$p(t|\psi) = |\langle t|\psi\rangle|^2, $
where the `time' kets $|t\rangle$ are related to the energy eigenkets $|E\rangle$ by the Fourier relation
$ \langle E|t\rangle = \gamma^{-1/2} e^{-iEt/\hbar}$
(where $\gamma$ is a suitable normalisation constant), analogous to the case of conjugate position and momentum observables.  This construction has recently been generalised to define canonical time observables for {\it any} quantum system having a discrete energy spectrum $\{E_j\}$, allowing the expectation value of any almost-periodic function of time to be calculated \cite{hall}.

In the final paragraph of \cite{hall} it was recalled that canonical time observables of the above type do not correspond to Hermitian operators acting on the Hilbert space of the system (for example, the time kets $|t\rangle$ and $|t'\rangle$ are not orthogonal for $t\neq t'$).  This strongly constrasts with a `time' observable previously proposed by Galapon, for a class of quantum systems having discrete energy spectra, which in the nondegenerate case corresponds to the Hermitian operator \cite{galapon}
\begin{equation} \label{gal}
T_G := i\hbar \sum_{j\neq k} (E_j-E_k)^{-1} |E_j\rangle \langle E_k| ,
\end{equation}
and satisfies $[T_G,H]=i\hbar$ on the subspace of states 
$S:=\{|\psi\rangle : \sum_j \langle E_j|\psi\rangle=0\}$ of the Hilbert space, where $H$ denotes the Hamiltonian operator of the system.

I claimed in \cite{hall} that the above subspace $S$ is not dense in the Hilbert space of the system. Galapon has pointed out, correctly, that this claim is wrong \cite{comm}.  I acknowledge and apologise for this error, which was primarily due to my use of an erroneous inequality (an embarrassing hybrid of the Schwarz and triangle inequalities that simply does not hold in general!). I further acknowledge that, as shown by Galapon \cite{galapon, comm}, the subspace $S$ is indeed dense.  One can even give explicit Cauchy sequences of states in $S$ which converge to the groundstate (or to any other energy eigenstate), eg, $\{\sum_{j=0}^N c_j(N)|E_j\rangle; N=1,2,\dots\}$, with
\[  c_0(N):= h(N)[\sigma(N)+h(N)^2]^{-1/2},~~~~c_j(N):= -(1/j)[\sigma(N)+h(N)^2]^{-1/2} \]
where $h(N) := \sum_{j=1}^N (1/j)$ and $\sigma(N) := \sum_{j=1}^N (1/j^2)$.
It is interesting to note that the denseness property is only valid for infinite-dimensional systems (as considered by Galapon) - if one formally defines $T_G$ as above for an $N$-dimensional Hilbert space, then the corresponding subspace $S_N$ is orthogonal to the vector $N^{-1/2}\sum_{j=1}^N |E_j\rangle$, and so is only $(N$$-$$1)$-dimensional.

However, problematic issues still remain for the interpretation of $T_G$ as a `time' operator.  For example, as (correctly!) pointed out in \cite{hall}, 
\begin{description}
\item[(i) ] The statistics of $T_G$ do not satisfy the covariance property (\ref{cov}), and so do not track the time evolution of the system (eg, $\langle T_G\rangle_\tau \neq \langle T_G\rangle_0 +\tau$).
\item[(ii)] The subspace $S$, on which the desired commutation relation
$[T_G,H] = i\hbar $
holds, is noninvariant (hence, if the relation holds at some time $t$, it will not in general hold at some later time $t'$).
\end{description}
Even so, one might ask whether $T_G$ might still provide useful `time' or `evolution' information of some sort (over short intervals, for example).  
Unfortunately, however, it appears that even for systems prepared in states restricted to $S$, there is no useful sense in which $T_G$ may be interpreted as a `time' operator.  In particular, $S$ itself has no nontrivial invariant subspace, and indeed
\begin{description}
\item[(iii)] A system can be described by an element of $S$ only at a set of times of total measure zero.
\end{description}
Hence, the conclusion in \cite{hall}, that ``{\it Galapon's operator, while well defined, has no clear connection with time at all}'', is maintained.

To demonstrate property (iii), define
$f(t):= \sum_j c_je^{-iE_jt/\hbar}$
for a given initial state $|\psi_0\rangle=\sum_j c_j |E_j\rangle$.  Hence, the state is a member of $S$ at time $t$ if and only if $|f(t)|=0$.  Suppose first that $f(t)$ is periodic, with period $\tau$.  Then, since the summation for $f(t)$ is one-sided (i.e., $f(t)$ is a causal function), property (iii) follows immediately from the corresponding Paley-Wiener condition \cite{paley} 
\[  \lim_{X\rightarrow\infty} X^{-1}\int_0^X dt~\left|\, \log |f(t)| \,\right| = \tau^{-1}\int_0^\tau dt~\left|\, \log |f(t)| \,\right| <\ \infty .  \]
More generally, $f(t)$ will be an almost-periodic function.  However, as per the  approximation theorem for such functions \cite{bohr}, there is a series of (causal) periodic functions which uniformly converges to $f(t)$, and the same result follows.

Finally, it is worth remarking that while the property $[T_G,H]|E_j\rangle =0$ claimed in \cite{hall}, is `flagrantly erroneous' as noted by Galapon \cite{comm} (since $|E_j\rangle$ does not belong the the commutator domain of $H$ and $T_G$), a related weak form of this property,
\begin{equation} \label{comm}  
\langle E_j | [T_G,H]|E_j\rangle = 0 \neq i\hbar ,
\end{equation}
does hold for all energy eigenstates. In particular, noting equation (\ref{gal}), the commutator $[T_G,H]$ may be {\it weakly} defined for a large class of states as
\[
[T_G,H] := -i\hbar \sum_{j\neq k} |E_j\rangle\langle E_k| = i\hbar\left( 1-|\chi\rangle\langle\chi| \right), \]
in terms of the (nonnormalisable) ket $|\chi\rangle:=\sum_j |E_j\rangle$ of the dual space, which immediately implies equation (\ref{comm}).

\end{document}